\documentclass[%
 reprint,
 amsmath,amssymb,
 aps,
]{revtex4-1}

\usepackage{graphicx}
\usepackage{dcolumn}
\usepackage{bm}

\begin{document}

\preprint{APS/123-QED}

\title{Spontaneous Arrangement of Two-way Flow in Water Bridge}

\author{Ping-Rui Tsai$^{1}$, Hong-Yue Huang$^{1}$, Cheng-Wei Lai$^{2}$, Yu-Ting Cheng$^{1}$, Cheng-En Tsai$^{2}$, Yi-Chun Lee$^{2}$, Hong Hao$^{1}$, and Tzay-Ming Hong$^{1\star}$}

\affiliation{%
$^1$Department of Physics, National Tsing Hua University, 
Hsinchu, Taiwan 30013, Republic of China}
\affiliation{%
$^2$Department of Chemistry, National Tsing Hua University, 
Hsinchu, Taiwan 30013, Republic of China}
\affiliation{%
$^3$National Hsinchu Senior High School,  Hsinchu, Taiwan  30013, Republic of China}%

\begin{abstract}
 By revisiting the century-old problem of water bridge, we demonstrate that it  is in fact dynamic and comprises of two coaxial water currents that carry different charges and flow in opposite directions. In the initial stage of setting up the water bridge, the inner flow is facilitated by the cone jet that is powered by H$^+$ and flows out of the positive-electrode beaker. A second and opposing cone jet from negative beaker is established {\it later} and forced to take the outer route. This spontaneous arrangement of two-way flow is revealed by using   fluorescein and carbon powder as tracers, and the Particle Image Velocimetry. These two opposing flows are found to carry non-equal flux that results in a net transport of water to the negative  beaker. A simple calculation based on energy conservation allows us to estimate the  flow speed and cross-sectional area of these co-axial flows as a function of time and applied voltage. 
 

\end{abstract}

\pacs{Valid PACS appear here}
\maketitle


Since 70.8 $\%$ of surface on earth is covered by sea, it is no wonder that scientists have shown great interest on the structure and properties of water, including extreme conditions such as water bridge (WB) at high electric field. This phenomenon \cite{WB} is realized in two beakers filled with deionized water and separated by a gap between 1$\sim$8 mm. After applying a DC voltage of about 1800 volt, a cone jet can be found to shoot from the positive beaker and establish a bridge across the gap after several attempts. This phenomenon was first reported in 1893 in a public lecture by the British engineer William Armstrong \cite{armstrong}. The fact that current remains less than 0.1 amp in spite of the high voltage  implies a very high resistance across WB. This is why WB often becomes
wiggly and eventually collapses in an hour due to heat \cite{WB}. Although there have been many hypotheses and experiments, the main mechanism behind WB and its structure remain contentious. 
For instance, although neutron scattering \cite{Neutron} and X-ray diffraction \cite{Xray} all failed to find any ordered structure and settled the debate that WB might present itself as a new form of water, the experimental group on Raman effect \cite{Raman} maintains that ``some changes in the scattering profiles after application of the electric field are shown to have a structural origin". Still more, the energy relaxation dynamics from infrared measurements\cite{IR} strongly indicate WB and bulk water differ at the molecular scale.

Why can WB hover in space?  
Fuchs \cite{WB} first ascribed this gravity-defying phenomenon to
electrostatic charges on the water surface due to the high electric field and high dielectric constant of pure water.
In 2009 Widom  {\it et al.}\cite{widom} provided a detailed calculation of the WB tension  in terms of the Maxwell pressure tensor in a dielectric fluid medium. In contrast, Gerald Pollack \cite{pollack} speculated that the bridge is made up of a hydronium lattice or Exclusion Zone water. 
By taking into account the charged nature of WB, Morawetz \cite{regive} has offered a theory to calculate not only the static and dynamic stability conditions, but also details such as the creeping height, the bridge radius and length, as well as the shape of WB.
Meanwhile, Skinner {\it et al.} \cite{Xray} proved in 2012 that there is no ordered structure in WB by high-energy x-ray diffraction experiments. They echoed the view by Aerov \cite{aerov} that the force supporting WB is the surface tension of water, while the role of electric field $\vec{E}$ is to avoid WB from breaking into separate drops.  We are doubtful of such a proposal because WB of length 3.86 mm can be achieved at $V=3000$ V. This is about four times longer than the maximum length of water column that can be sustained by the surface tension at $V=0$. However, an electric field  of $E=10^6$ V/m can only increase the surface tension coefficient by $6\%$ \cite{surfacetension} that may be far too weak to support WB.

\begin{figure}
\centering
\includegraphics[height=5.7cm,width=8.5cm]{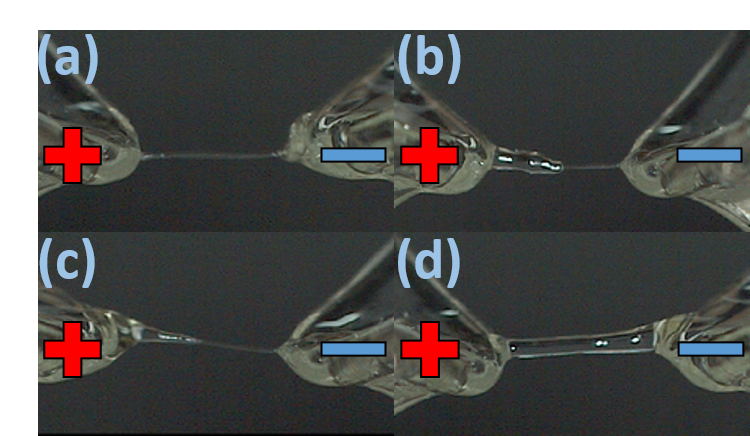}\caption{(color online)  (a) Cone jet shoots from the positive beaker and arrives at the negative one after several unsuccessful attempts. (b, c) Negative cone jet takes advantage of the connection in (a) and flows from negative  to positive beakers by climbing on its surface. Like shovelling  snow,  it scrapes  and causes water to pile up temporarily on the left side. (d) A thicker and more uniform WB is eventually stabilized. Videos can be found in SM I-1 of Supplemental Material (SM)\cite{sm}. } 
\label{2}
\end{figure}

Let's save the debate on the origin of WB  for later discussions and first concentrate on exploring the possibility that a more macroscopic and yet ordered structure than the debunked crystalline one may exist. We believe the fact that the radius of H$^+$ ions is much smaller than that of OH$^-$ renders the front edge of cone jet from the positive beaker being crammed with more ions and thus enjoying a stronger propulsion. By using a high-speed camera, we observe that, as soon as this cone jet \cite{cone1,cone2} succeeds at landing at the negative beaker as in Fig. 1(a), a surge of counter flow  is observed  in Fig. 1(b,c) to climb from the latter and advance on the surface of this newly-built connection. This stacking process continues until the cross section of WB stabilizes in Fig. 1(d).

One immediate question is whether this spontaneous separation of opposing flows remains true after WB is fully stabilized. To answer this question, tracers become useful at helping us visualize the flow field. Two cautions worth noting here. First, WB is sensitive and liable to  ions that may accompany the addition of indicators. Second, avoid inserting the pH meter into the beaker because, like any other contact instruments, such as multimeter, electronic thermometer, and microfluidic device, it will likely crash under the high voltage. Fluorescein \cite{fluo} is the first indicator we adopt to trace the direction of internal flow. Its solubility in water is merely 50 mg/L  with pKa = 8.72, indicating that the amount of ions generated by this compound is negligible. When added to the positive beaker,  fluorescein can be seen to flow through the inner layer of WB in Fig. 2(a) and create a trail of fluorescence in the negative beaker. In contrast, the tracer occupies only the outer layer of WB when added to the negative beaker in Fig. 2(b). 


\begin{figure}
\centering
\includegraphics[height=9.5cm,width=6.5cm]{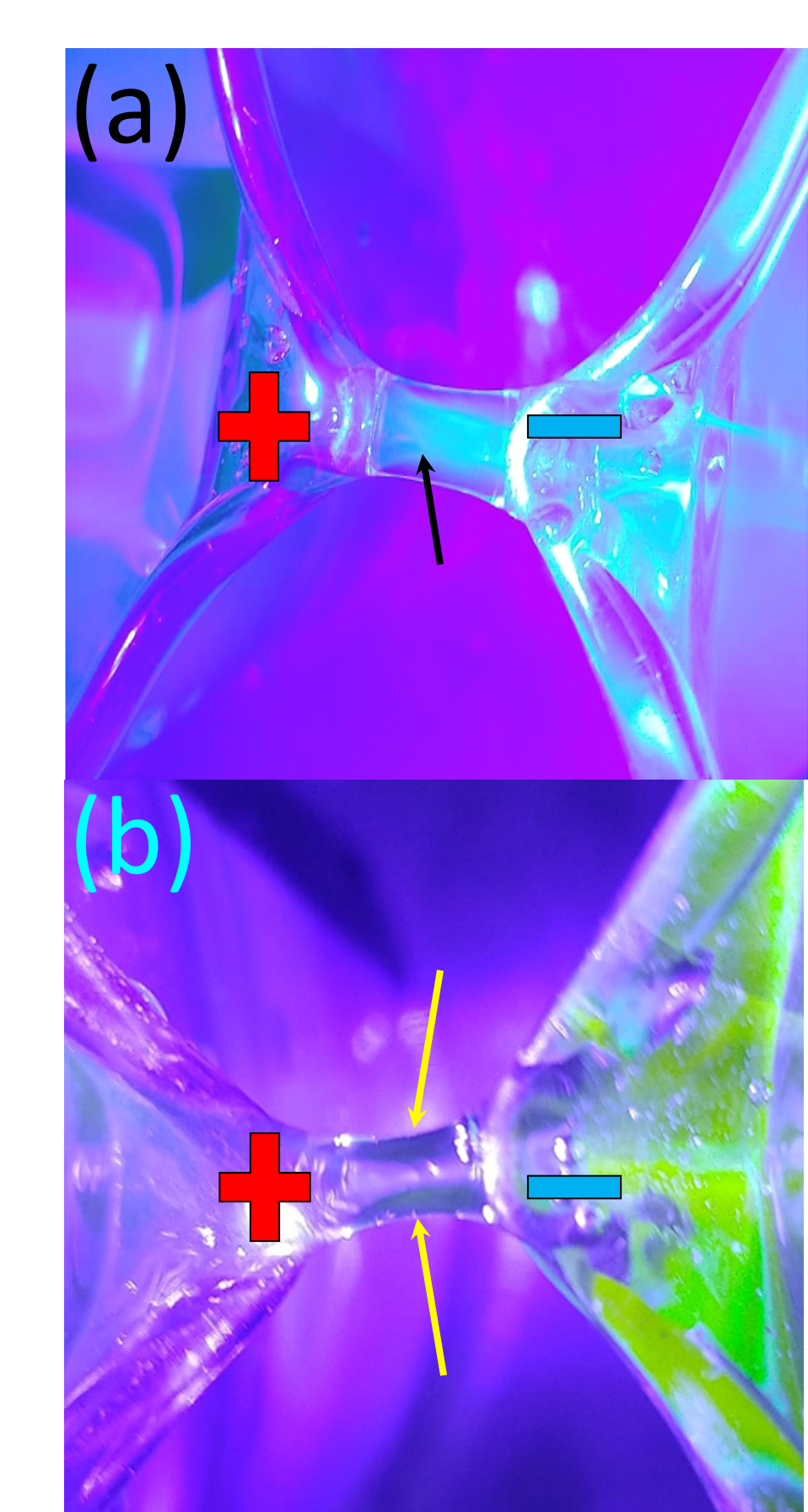}\caption{(color online)  Fluorescein is used to determined the direction of inner flow for WB at  10000V. 
A blue laser of 1W and 450nm is used to illuminate the tracer added to the positive  and negative beakers in (a) and (b), respectively.  The trail of fluorescein  all runs toward the opposite beaker, but takes the route of inner and outer layers of WB, separately. More details can be found in SM I-2 and SM 3  \cite{sm}. }
\label{2}
\end{figure}

As a surface tracer, carbon powders are tested  in SM 1-1\cite{sm} by the capillary electrophoresis \cite{carbon_neg1} to be neutral in charge  before being dispersed on either beaker in Fig. 3(a). 
These powders are observed to travel only from negative  to positive beakers. Because they always float on the water, this corroborates the picture of two-way flow  established by fluorescein. According to  SMI-6  \cite{sm}, we measured the pH value of water and found the positive beaker becomes more alkaline with time, while the negative beaker more acidic. This leads us to conclude that the opposing flows from negative  and positive beakers must be powered by OH$^-$ and H$^+$ ions and take the route of outer and inner layers of WB.

\begin{figure}
\centering
\includegraphics[height=7cm,width=8.7cm]{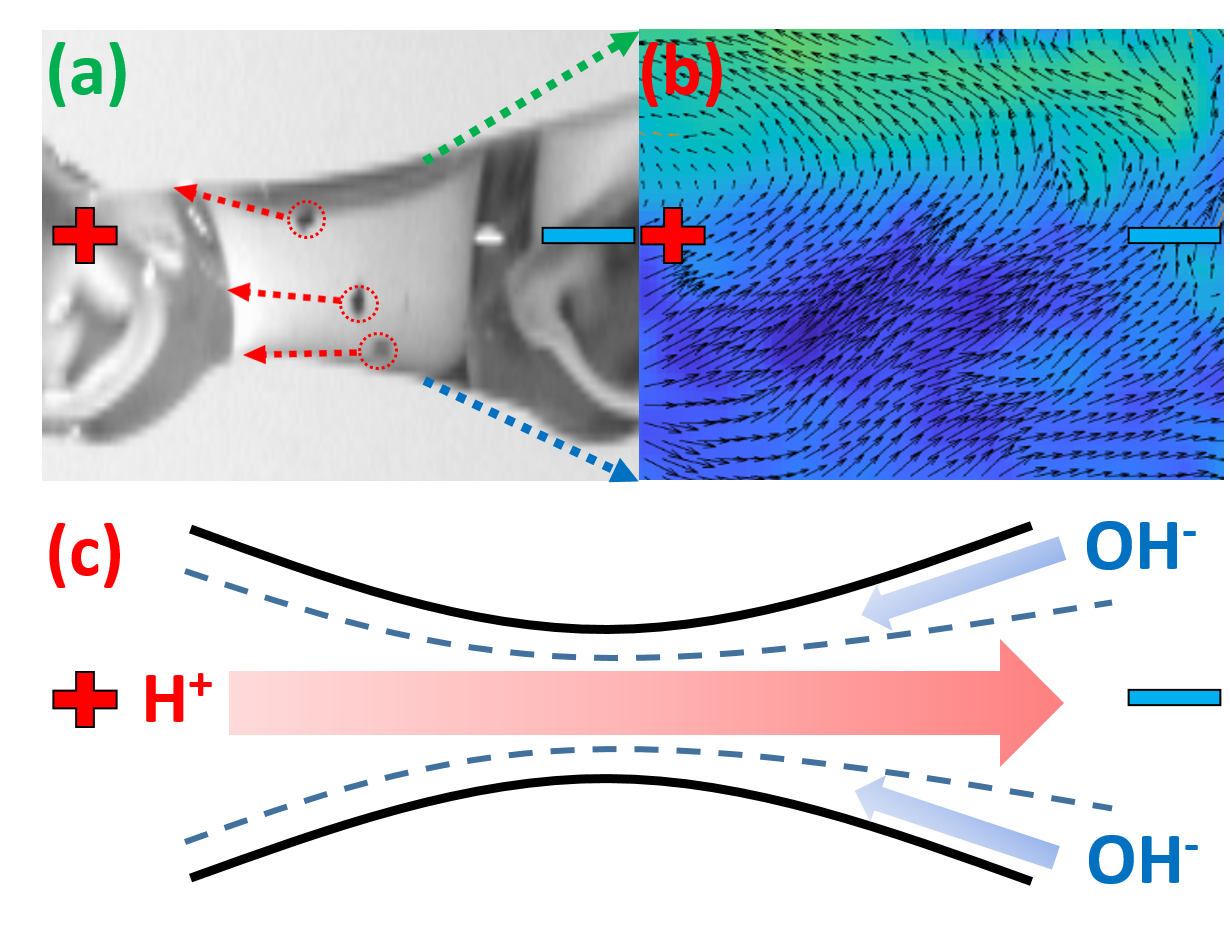}\caption{(color online) Red arrows in (a) indicate that  carbon powers only move from negative to positive beakers. Because they float on  water, this confirms the unidirectional flow  of the outer layer. The   PIV image in (b) corroborates that water in the outer/inner layer on the green/blue background moves towards the positive/negative beakers. Schematic plot in (c) summaries the arrangement of two-way flow. Clearer videos and pictures can be found in  SM 1-2, SM 2-2, SM 2-3, and SM I-3 \cite{sm}.}
\label{4}
\end{figure}
In order to visualize the flow field in more detail, we appeal to Particle Image Velocimetry (PIV) that is composed of one concave lens and one cylindrical lens. Laser of 1W and 450nm passes through the optical path and illuminates the PVC particles \cite{PIVparticle} inside WB. The reflected light signal is picked up by a high speed camera with 1000$\sim$4000 fps. We perform image processing to enhance particle signals and do denoising by the Gaussian filter with the standard error 1.1. Then, Sobel edge detection is employed to determine the profile. Finally we use the  Morphology to dilate the PIV image to make the particle position more discernible, as shown in SM part 3. Afterwards, pre-processing datas are converted to images via the MATLAB toolbox, PIVlab\cite{PIVtool}, for the analysis of flow direction\cite{PIVE1}. A sample image is shown in Fig. 3(b) that reveals the flow vectors in the inner layer of WB mostly point towards the negative beaker. Some vortexes inevitably occurred  due to collisions with the opposite flow on the outer layer. To be sure, we have checked in SM 2-1 \cite{sm} that PIV particles of PVC are not affected by the high electric field. Figure 3(c) summarizes our findings of two-way flow so far. More images that were taken at different horizontal cross-sections can be found in part 3 of SM \cite{sm}.
\begin{figure}

\centering
\includegraphics[height=7.9cm,width=8.7cm]{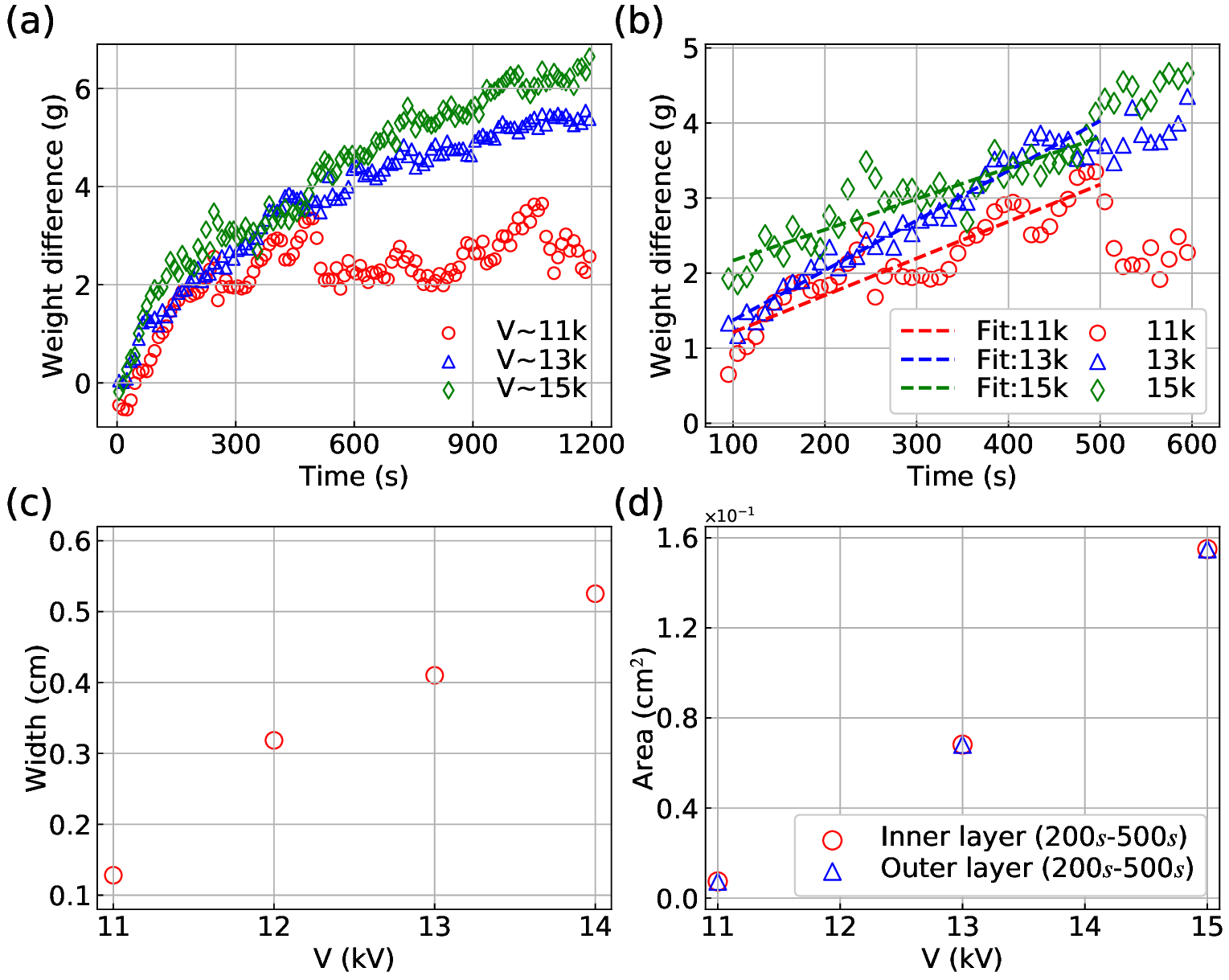}
\caption{(color online) The weight difference between negative and positive beakers is recorded in (a) as a function of time for three voltages. Plot (b) shows that data in (a)  can be fit by a linear line during the initial 100$\sim$500 sec. The reason for using this time frame for our theoretical estimation is that WB is generally unstable  before 100 sec and to avoid the complication of  pressure difference built up by the principle of connecting pipes after 500 sec. In (c), the width of WB is experimentally found to increase with voltage, consistent with the trend of cross-sectional area for the inner and outer layers in  (d) estimated by inputting  (a) into Eq. (1).  }

 \label{2}
\end{figure}

It has been reported \cite{moto} that the water level of positive beaker will fall below  the negative one during the WB experiment. The growth rate of their weight difference decreases with time in Fig. 4(a, b), presumably due to the counter flow from the buildup of pressure difference in connecting pipes.  
The net flux of water in WB can be calculated by
\begin{equation}
    I(V,t)=m\rho_{H_{2}O}\Big[A_{+}(V,t)v^{+}_{H_{2}O}-A_{-}(V,t) v^{-}_{H_{2}O}\Big]
\end{equation}
where $m$ and $\rho_{H_{2}O}$  are the molecular mass and  number density of water, $A_{\pm}$ denotes the cross-sectional area in inner and outer layers, and $v^{\pm}_{H_{2}O}$ are their corresponding flow speed for water \cite{notation}.
 A quick way to estimate $v^{\pm}_{H_{2}O}$ is via energy conservation, namely, the kinetic energy of flowing water is converted from the electrostatic energy of ion:
\begin{equation}
    v^{\pm}_{H_{2}O}=\sqrt{\frac{2eV\rho_{\rm ion}}{\epsilon m\rho_{H_{2}O}}}
\end{equation}
where $e=1.6\times 10^{-19}$ C, the dielectric constant  of liquid water is $\epsilon=78.4$, and the number density $\rho_{\rm ion}$ of ions can be approximated by pH=7.  By tracing the movement of carbon powers, we  estimated $v^{-}_{H_{2}O}$ in the outer layer of WB with length 0.1 cm and V=11000 $\sim$ 15000 V to be around  0.11  $\sim$ 0.18 m/s, about an order of magnitude smaller than the predicted value of $1.62\sim 1.8
9$ m/s from Eq. (2). We believe this discrepancy comes from the creation of vortexes and the fluctuation and non-uniformity for the total cross-sectional area $A_{\rm total}\equiv A_{+}+A_{-}$ of WB. Due to the small dimensions of WB, the effect of these unavoidable inelastic processes becomes more pronounced and renders the conversion rate from the electrostatic energy  to  propelling the water flow very poor - less than $\sim$1$\%$ .    

How about the energy lost to heating? For a volume of water  of $3.14\times 10^{-9}\ {\rm m}^{3}$ that makes up WB, the thermal loss is estimated to be about 0.131 joul for the initial 100 $\sim$ 500 sec during which the temperature was raised by roughly $10^{\circ}$K. This is negligible compared to the energy input of  4000 joul from the power supply for a current of $1 $ mA and voltage $10000$ V in the same period. 

Empirically the net flux can be read off from Figs. 4(a, b) as half of their derivative. By plugging Eq.(2) into (1) and using Figs. 4(c) as an input information for  $A_{\rm total}$, we could estimate $A_+$ and $A_-$. As shown in Fig. 4(d), these two cross-sections increase with voltage and are comparable: ${A_{+}}/{A_{-}}\sim 1+2.00\times 10^{-3}$, $1+2.74\times10^{-4}$ and $1+6.97\times10^{-5}$ for 11000, 13000, and 15000V. 

Before concluding this work, allow us to mention another interesting experiment of ours that may contribute to clarifying the origin of WB.    
We are in favor of the more conventional view \cite{marin} that, since there is no bridge-like structure if water is replaced by nonpolar liquid such as n-hexane or ethanol, the potential energy $-{\vec p}\cdot {\vec E}$ between water dipoles $\vec{p}$ and $\vec E$ must play an important role.  
To verify this scenario, we destabilize WB by dripping extra water onto it by a burette.
 The  excessive water is observed to hang like a drool at the bottom of WB, and eventually detach by a pinch-off at the bottle-neck, as shown in Fig. 5(a). The remaining part of WB bounces back to a quasi-static height $h$ before embarking on a much slower process of reducing back to its original shape. The relation between $h$ and voltage is shown in Fig. 5(b). For comparison, $h$ is strictly zero for the water column at $V=0$ in Fig. 5(b) that is maintained solely by the surface tension. A back-of-the-envelop calculation using $mg\approx \nabla\left(\vec{p}\cdot \vec{E}\right)$ gives
\begin{equation}
    h=\sqrt{\frac{Vp}{\epsilon mg}}.
\end{equation}
 This readily captures the concave and increasing trend of Fig. 5(c) and predicts  the right magnitude for $h$.

\begin{figure}
\centering
\includegraphics[height=5cm,width=8.7cm]{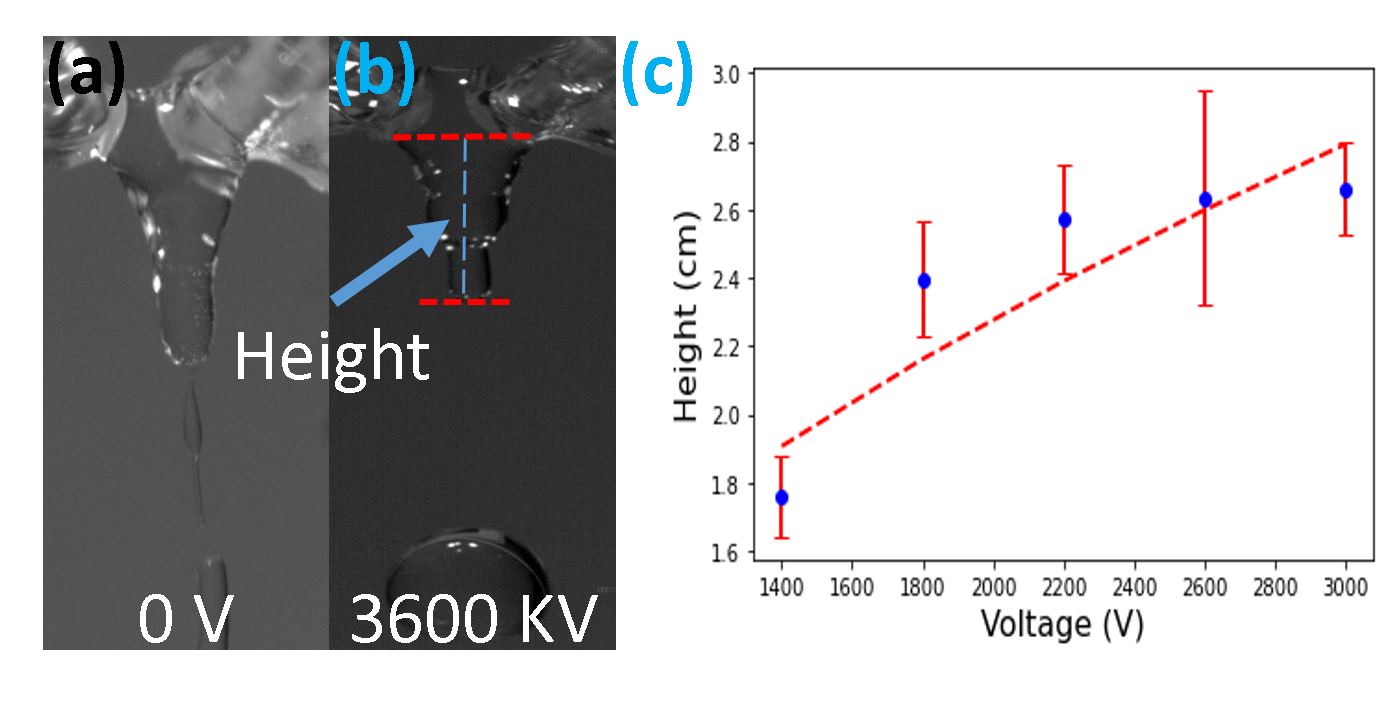}
\caption{(color online) In contrast to the 0.1cm-long WB  at 1800 V in panel (b), the 0.05-cm water column in (a) was  realized by separating two initially joined beakers at zero voltage. When extra water was poured onto both bridges, case (a) would shed off the water by one fell swoop, while (b) dividing it into several drops.  Panel (c) shows how the drool height $h$ in (b) changes with voltage. Note that case (a) does not exhibit such a quasi-static drool. More details can be found in SM I-5a, SM I-5b, and SM 4 \cite{sm}.}
 \label{2}
\end{figure}

In conclusion, the dynamic structure of WB is rigorously proven by experiments to comprise of two-way flow. Although this spontaneous separation into multiple layers is not new to fluid dynamics, e.g., kuroshio \cite{kuroshio} due to different salinity and temperature, the spatial arrangement in WB is special at that it occurs in the millimeter scale. Again, unlike the example of liquid crystals\cite{CRY1,CRY2}, WB involves two milli-flows that go in totally opposite directions. Microscopically water molecules in the inner layer  are powered by H$^+$ ions that move under the high voltage, occupies a slightly bigger cross-sectional area in WB, and flows from the positive  to negative beakers. In contrast, OH$^-$ ions drive  the flow in the outer layer flows towards the opposite direction. 
Note that, although Ref. \cite{regive} has theoretically studied the net flux of water in WB, the author assumed the flow to be uni-directional and that the surface flow was negligible. Both assumptions contradict our observations and estimation of flow speed  10 cm/s at the outer layer from the carbon-powder experiment.

Our research did not rule out the possibility that H$_{3}$O$^{+}$\cite{H3O1,H3O2,H3O3} may exist in WB. Researchers along this line of thought may want to focus on the inner flux. It will be interesting to test whether this structure of two-way flow and the imbalance between opposing fluxes also exist in  other polar liquids, such as glycerin \cite{Glycerin1}.  In the mean time, we suggest that a re-examination of data from x-ray diffraction and Raman spectroscopy by taking into account this complex and dynamic structure deduced by our research. Furthermore, it will be desirable to employ the technique of small-angle X-ray scattering \cite{smallangle1,smallangle2} on the outer layer of WB to explore possible difference in the electronic density. A potential application of this unique arrangement of two-way flow  may be in the simulation of action potential of neurons that has so far relied on the equivalent circuits\cite{NEU1} and recent interest at studying liquid flow and control in miniaturised fluidic circuitry without solid walls\cite{nature}.

We benefit from useful discussions with Jow-Tsong Shy, Sun-Ting Tsai, Li-Min Wang, and Hung-Chieh Fan Chiang, and technical supports from Yi-Shun Chang, Chun-Liang Hsieh, Zhen-Man Tian, Yen-Wen Lu and Chin-Fa Hung. Financial support by MoST in Taiwan  under grants 105-2112-M007-008-MY3 and 108-2112-M007-011-MY3 is also gratefully acknowledged.


\begin{thebibliography}{}
\bibitem{notation} A list of notation for all parameters in our derivations can be found in SM \cite{sm} (SM I and SM 1-1, SM1-2, SM 2-1, SM 2-2, SM 2-3 and SM 2-4).
\end{thebibliography}
\end{document}